\documentclass[a4paper]{VisionStyle}
\usepackage{epsfig}

\def\exosat{{\it EXOSAT~\/}}
\def\ginga{{\it Ginga~\/}}
\def\rosat{{\it ROSAT~\/}}
\def\asca{{\it ASCA~\/}}

\def\xmm{{\it XMM-Newton~\/}}

\def\chandra{{\it Chandra~\/}}
\def\sax{{\it BeppoSAX~\/}}

\def\H0{{\rm ~km~s^{-1}~Mpc^{-1}}}

\def\norm{{\rm ~photon~cm^{-2}~s^{-1}~keV^{-1}}}

\def\et{et al.~\/}
\def\eg{{\it e.g.~\/}}

\def\ie{{\it i.e.~\/}}

\def\la{\mathrel{\hbox{\rlap{\hbox{\lower4pt\hbox{$\sim$}}}{\raise2pt\hbox{$<$}}}}}
\def\ga{\mathrel{\hbox{\rlap{\hbox{\lower4pt\hbox{$\sim$}}}{\raise2pt\hbox{$>$}}}}}

\def\d25{D$_{25}$}

\def\los{line-of-sight\thinspace}
\def\.25{0.25 keV\thinspace}

\begin{document}

\title{What can we learn from EPIC X-ray spectra of Seyfert 1 galaxies?}

\author{K.\,A. Pounds and J.\,N. Reeves} 

\institute{
Department of Physics and Astronomy, University of Leicester,
University Road, Leicester, LE1 7RH }

\maketitle 

\begin{abstract}

The EPIC detectors on \xmm provide the most sensitive broad band
($\sim 0.3-12$~keV) X-ray spectra to date. Observations of 6 Seyfert
1 galaxies, covering a wide luminosity range, are examined with the aim of 
identifying the primary X-ray continuum and constraining
superimposed emission and absorption features.
A soft excess $\it emission$ component is seen in every case, but with a 
spectral form 
which differs 
markedly with luminosity across our sample. Current interpretations of the 
soft excess range 
from intrinsic thermal emission from the accretion disc 
to reprocessing of harder
radiation absorbed in the disc skin. Visual examination of the
broad-band EPIC spectra suggest that
the luminosity trend in the observed spectral profiles may be governed 
primarily by differences in the line-of-sight absorption. 
In that case the underlying continuum could have a common form across
the sample. Examination of spectral features in the Fe K band confirm
the common presence of a narrow emission line at $\sim$ 6.4~keV.
Modeling of the EPIC
spectra above $\sim$ 7~keV is shown to be critical to quantifying (or
confirming) a broad Fe K line in at least some cases.

\end{abstract}

\section{Introduction}

For more than a decade after powerful X-ray emission was established
as a common property of AGN (\cite{kpounds-C2:elvis78}, Pounds 1977)
little insight on the nature of the X-ray source was gained from
spectra which were well-modelled by a featureless power law continuum
of photon index $\Gamma$~$\sim1.5-2$. The increased spectral bandwidth
of the combined ME and LE detectors on \exosat yielded evidence that
the spectrum of many AGN steepened below $\sim0.2-12$~keV, the
so-called 'soft excess' (Arnaud 1985, Turner \& Pounds 1989). Since
that time new X-ray satellites have brought major improvements in
spectral resolution, sensitivity and bandwidth, providing X-ray
spectra of increasing complexity (and diagnostic potential). The
particular strengths of each new mission has led to new areas of
research, with the high energy response of \ginga establishing the
widespread importance of `reflection' (Nandra \& Pounds 1994),
\rosat showing absorption in \los
photoionised matter to be surprisingly strong (Turner \et 1993), 
\asca resolving discrete
spectral features, including a broad fluorescent iron K line (Tanaka
\et 1995, Nandra \et 1997) and \sax
providing uniquely broad bandwidth data. 

\chandra and \xmm have now taking this evolution a large step further, with
high-throughput grating spectra providing a qualitative
improvement in spectroscopic detail. The most striking results to date
have been in the detection of complex absorption and emission spectra
in the soft
X-ray band (e.g. Kaspi \et 2001) where the 1- or 2-stage photoionisation
model of the `warm absorber' (e.g. Reynolds 1997) has been shown to be
an inadequate description. Instead, the \xmm and
\chandra grating spectra have shown the (often outflowing) material
illuminated by the central X-ray source to exhibit a remarkably wide range
of ionisation stages (Behar \et 2001). Interpretation and modelling of 
these data is already promising
a dramatic improvement in mapping the structure, content and
dynamics of matter in the nuclear region of
Seyfert galaxies. Conversely, modelling the temperature and ionisation
of
this gas will provide a unique measure of the total ionising flux,
including the often dominant XUV component invisible from Earth. 

To date, less progress has been made in utilising the 
diagnostic
potential of the broader-band spectra from the EPIC cameras on \xmm;
to briefly explore that potential is the purpose of the present paper 
which brings together EPIC
spectra from 6 Seyfert galaxies observed early in the mission and 
covering the (2-10 keV) X-ray luminosity range
$10^{43}-10^{45}$erg sec$^{-1}$.   

\section{The Seyfert sample}

The present sample represents the start of an ambition to develop a
'big picture' of the X-ray spectra of radio-quiet AGN, to identify
general features and trends, and attempt to determine the 
influence of the `primary 
variants', namely the luminosity, relative 
accretion rate and black hole mass. Table 1 lists the sources included
in the present sample, in order of X-ray 
luminosity, together
with the dates, exposure time and original status of each observation.
Figure 1 shows the broad spectral features for these AGN,
in each case plotted 
as a ratio of
the observed EPIC spectrum to a simple power law plus Galactic 
absorption model. Data from the PN camera are used in all the figures
but the analysis and interpretation that follows is also consistent with
data from the MOS cameras.
For an initial visual
comparison all spectra have been fitted in the 2--12 keV band and
then extrapolated to the lower limit of the reliably calibrated EPIC data.

Several common features stand out, together with some clear
differences and apparent trends. These are discussed in the following
sections. 

\begin{table}
\centering
\begin{minipage}{85 mm} 
\caption{Observing log of EPIC-PN spectra}
\centering
\begin{tabular}{lcccc}
\hline
source & date & exp (ks) & Lx$^{a}$ & $\Gamma$$^{b}$\\\hline
MCG-6-30-15 & 11-07-00 & 53.8 & 0.6 & 1.69\\
Mkn766 & 20-05-00 & 35.6 & 0.7 &1.97\\
NGC5548 & 9-07-01 & 30 & 3.8 &1.61\\
NGC5548 & 12-07-01 & 46 & 5 & 1.70\\
Mkn509 & 25-10-00 & 27 & 14 & 1.64\\
Mkn509 & 20-04-01 & 30.5 & 22 & 1.74\\
1H0419-577 & 4-12-00 & 5.8 & 57 & 1.88\\
PKS0558-504 & 24-05-00 & 10.2 & 79 &1.88\\

\hline
\end{tabular}
\label{label2}
\footnotetext[1]{X-ray luminosity ($10^{43}$erg s$^{-1}$) 
at 2--10keV; $H_{0} = 50$}
\footnotetext[2]{power law index 
at 2--10keV; 3--10keV for PKS0558-504}

\end{minipage}
\end{table}

\section{The Soft Excess}

As noted earlier, the `soft excess' was first claimed to be a common
feature in AGN X-ray spectra from \exosat observations of Seyfert galaxies 
(Turner \& Pounds
1989). Subsequent \rosat observations showed the spectra of AGN were
often significantly steeper than $\Gamma$~$\sim2$ below $\sim1$~keV, 
although the evidence of absorption features led to some confusion in 
the literature between soft spectral
{\it slope} and excess soft {\it emission}. \sax and now EPIC spectra 
have removed
such
ambiguities and
confirm that a 
soft X-ray {\it emission}
component is indeed a common feature
of Seyfert 1 galaxies, as Figure 1 clearly shows.

\begin{figure*}
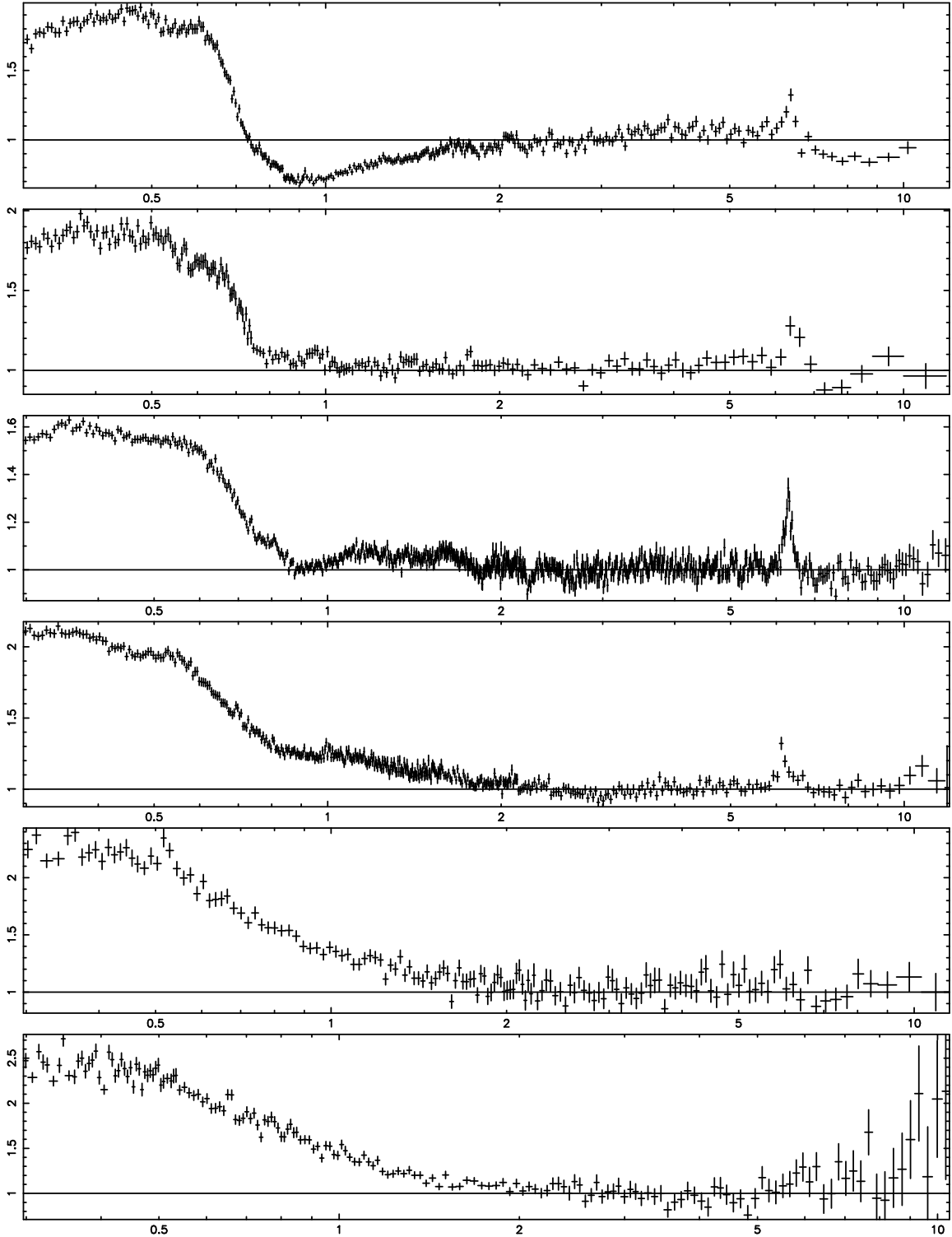
 
\centering
\vbox{
\includegraphics[width=3.3 cm, angle=270]{kpounds-C2_fig1a.ps}
\includegraphics[width=3.3 cm, angle=270]{kpounds-C2_fig1b.ps}
\includegraphics[width=3.3 cm, angle=270]{kpounds-C2_fig1c.ps}
\includegraphics[width=3.3 cm, angle=270]{kpounds-C2_fig1d.ps}
\includegraphics[width=3.3 cm, angle=270]{kpounds-C2_fig1e.ps}
\includegraphics[width=3.3 cm, angle=270]{kpounds-C2_fig1f.ps}
}
\caption{EPIC PN camera spectra for 6 Seyfert 1 galaxies shown as the
ratio of measured flux to the 2--12 keV powerlaw fit as listed in Table 1. 
From the top, in
order of increasing 2--10 keV luminosity, are MCG-6-30-15, Mkn 766,
NGC 5548, Mkn 509, 1H 0419-577 and PKS 0558-504. The abissca is
photon energy (keV) in the observer's frame.}
\end{figure*}

Interpretation of the soft X-ray emission remains uncertain, however,
and
indeed may differ across the present small sample of Seyfert galaxies. 
In general terms the alternatives
are primary emission from the accretion disc, representing
gravitational energy released by viscosity in the disc, or
secondary radiation from the re-processing of hard X-rays 
in the surface layers of the disc. Clarifying these alternatives is 
critical to determining where the primary energy release occurs  
and thereby to understanding better the fundamental mechanism(s) by
which AGN produce such powerful X-ray emission.
Broad-band observations, particularly from {\it Beppo-SAX}, 
have shown a simple power law
to approximately describe the hard continuum spectra of several Seyfert 
galaxies 
over two decades in energy, supporting an origin by multiple scattering of low
energy (disc) photons by high energy (coronal) electrons. However, it
remains unclear, in general, whether the primary (gravitational)
energy is released directly into the corona (eg by magnetic field
reconnection) or as thermal emission from the accretion disc. In
either case the re-absorption of 
hard coronal X-rays in the disc surface, as in the
`disc-corona' model (\eg Haardt \& Maraschi 1991), 
offers an important feedback mechanism to explain the similarity of
power law slopes in many AGN. Attempts to clarify the
`driver' in this process by determining `lags' or `leads' between
spectral
components, have yielded no clear picture. Potentially the most direct 
indicator
- the energy balance of the spectral components - is likely to remain out of
reach, since the intrinsic thermal emission
from the accretion disc will peak in the (unseen) XUV band
for AGN with central black holes masses greater than $\sim10^{6}~M_{\odot}$.

Visual examination of the present EPIC spectra offers some interesting
insights. For the lower luminosity sources in our sample (MCG-6-30-15,
Mkn 766, and possibly NGC 5548) the soft excess rises very sharply below
$\sim0.7$~keV. This `sharp soft excess' (SSX hereafter) is indicative of 
an atomic
feature, which - combined with the need for {\it emission} above the power
law extrapolation - is consistent with an interpretation as hard X-rays 
re-processed  
in the inner disc. Such an explanation has been made for MCG-6-30-15 and
Mkn 766,
with RGS spectra apparently showing strong relativistically
broadened recombination lines of OVIII, NVII and CVI
(Branduardi-Raymont \et 2001, Sako \et 2001). 

The EPIC soft X-ray spectra of Mkn 509, 
1H 0419-577, and PKS 0558-504 are of quite different appearance, with a
gradual upward curvature emerging below $\sim3$~keV. In the
latter 2 sources this gradual soft excess (hereafter GSX) has been interpreted
as Comptonised thermal disc emission (Page
\et 2002, O'Brien \et 2001). The absence of strong 
discrete
spectral features in the corresponding RGS spectra of Mkn 509 and PKS
0558-504 lends important support for this view. (The observation of 
1H 0419-577 unfortunately yielded only
PN camera data).

The recent comparison of the EPIC spectrum of 1H 0419-577 with
historical data from {\it ROSAT}, \asca and \sax (Page \et 2002) provides 
some of the best evidence to date for the soft X-ray
emission being the `driver' of spectral change, with 1H~0419-577
apparently switching over several years between states when the soft 
X-ray excess was weak
or absent, and the more typical bright state observed by \xmm
(Figure 2). 
Significantly
the power law component was much harder when the soft flux was weak,
implying a photon-starved corona (in the context of the disc-corona
model).

NGC 5548 is of intermediate luminosity in the present sample, and has a
soft excess whose form lies between those of the above groups. As such it may
hold the key to linking the disparate interpretations of the low and
high luminosity sources outlined above. The \chandra
spectra of NGC 5548 
are reported to
show many strong absorption lines (Kaastra \et 2000), 
and further
refinement of the instrument calibration may show that
the structure evident in the EPIC ratio plot below $\sim3$~keV 
(Figure 1)
is associated with absorption by ionised matter. Such considerations
lead to the
speculation that the principal cause of the changing shape of the soft
excess across the sample is absorption, affecting the broad-band
spectrum from $\sim0.7$~keV up to $\sim3$~keV (e.g. see Nicastro \et 1999).
If that is so, the observed 
trend with luminosity would 
have a natural explanation in terms of the increasing ionisation (and
low energy transparency) of a substantial column of absorbing gas, 
with increasing
luminosity.

An implication of this interpretation is that until the total absorption 
can be modelled in sources
such as Mkn 766 and NGC 5548, the strength (and physical nature) of 
the soft X-ray emission will remain uncertain. Progress will require
the simultaneous analysis of well-calibrated spectra from
dispersive instruments, such as the \xmm RGS, and broad-band detectors,
like EPIC, in order to quantify both line and edge absorption over the
observed energy range.   

\begin{figure}
\centering
\begin{minipage}{85 mm} 
\centering
\hbox{
\includegraphics[width=8 cm, angle=0]{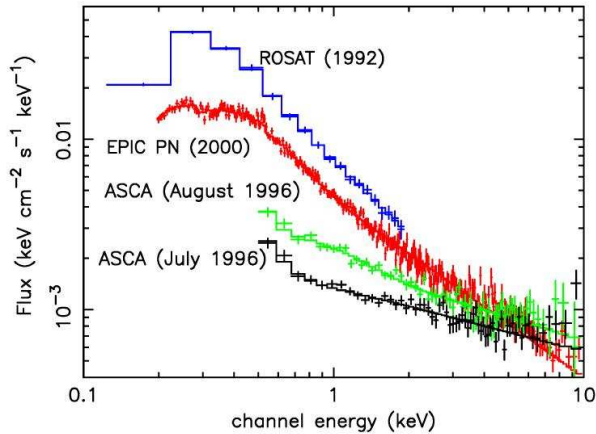}
}
\caption{Unfolded spectra showing the dramatic change in state of 1H
0419-577 over the period 1992-2000}
\end{minipage}
\end{figure}

\section{Fe K emission line}

\subsection{Narrow Fe K line}

A narrow emission line at $\sim$6.4~keV, with equivalent width in the range
50--100 eV is emerging as a common component of Seyfert 1 galaxies (at
least for a 2--10 keV luminosity less than $10^{45}$ erg sec$^{-1}$).
Figure 3 shows the EPIC spectral ratio for NGC 5548 where the narrow
line is the most obvious spectral feature in the 2--12 keV band. The
interpretation of this line as Fe K fluorescence from cold matter
distant from the inner disc region seems clear. Reflection from
the putative molecular torus has been suggested by several authors
and remains a strong candidate, the apparent weakening of the line for 
very luminous sources providing
circumstantial support in the context of the torus being flattened by
radiation pressure.
However, it remains possible that a significant fraction of the narrow line is
produced at much smaller distances from the central continuum source, 
with the outer disc and broad line clouds  
being realistic candidates. A further possibility might be by scattering in a 
population of 'clouds' sufficiently dense to survive at small
radii (Guilbert \& Rees 1988) and perhaps indicated in the recent
detections
of strong Fe K absorption edges in several \xmm spectra (Boller \et
2001, Reeves \et 2002).
Just as the broad Fe K line has been heralded as a key diagnostic of
strong gravity near the putative black hole, the narrow Fe K line may
prove to be a valuable tool for mapping dense matter throughout the
nuclear region of AGN. 
The route to constraining the origin of the narrow Fe K line will be 
from improved measurements of the line width and flux variability. A recent 
\chandra measurement
has resolved the line in NGC 5548 with FWHM $\sim$4500~km~s$^{-1}$ (Yaqoob
\et 2001) suggesting
an origin within the BLR. Figure 3 illustrates the narrow
line
in \xmm data 
taken $\sim3$ days apart, compared in each plot with a simple power
fit over the $\sim3-10$~keV band. Intriguingly, the relative intensity
of the line appears very similar, while the 2--10 keV flux increased 
by 40 \%
between the 2 observations. However, the increase above 7 keV (\ie
in the flux capable of causing the fluorescence) the increase was only 20 \%,
rendering inconclusive the measured 'increase' in the narrow Fe K line 
intensity,
from 2.5 (2.0--3.1) to 3.0 (2.3--3.9) $\times10^{-5}$ $\norm$.

\begin{figure}
\centering
\begin{minipage}{85 mm} 
\centering
\hbox{
\includegraphics[width=6 cm, angle=270]{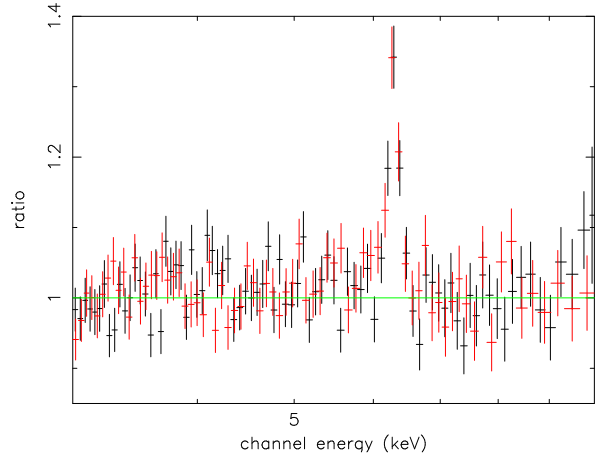}
}
\caption{Comparison of the narrow Fe-K line in observations of NGC 5548
in successive XMM orbits. The relative strength of the line appears
unchanged though the underlying 2--10 keV continuum has increased by a factor
$\sim1.4$, but see the text}
\end{minipage}
\end{figure}

\begin{table*}
\begin{center}
\caption{Fit parameters to the 0.5--10 keV spectrum of MCG-6-30-15 
($\chi^{2}$/dof$\sim$1304/1189)}
\begin{tabular}{p{3.0truecm}p{0.2truecm}p{0.01truecm}p{1.5truecm}p{1.5truecm}
p{2.0truecm}p{0.2truecm}p{0.01truecm}p{1.0truecm}}
\\\hline
continuum & $\Gamma$ &  $\sim$ & 1.86 && absorption & E& $\sim$ & 7.3~keV\\
	  & kT$_{1}$& $\sim$& 100~eV & & edge & $\tau$& $\sim$& 0.1\\
          & kT$_{2}$& $\sim$ & 210~eV \\
&\\
narrow line & E & $\sim$ & 6.38~keV &&broad (Laor) & E & $\sim$ & 6.4~keV\\
&$\sigma$& $\sim$ & 0.1~keV && line  &$\beta$ & $\sim$& 5\\
& EW & $\sim$& 90~eV & & &  r$_{in}$ & $\sim$ & 1.25~r$_{g}$\\
 & & & & &	  & r$_{out}$ & $\sim$ & 400~r$_{g}$\\
warm absorber & $\xi$ & $\sim$ & 5.9 & &	  & $\theta$ & $\sim$
& 30$^{o}$ \\
 & $N_{H}$ & $\sim$ & $10^{22}$cm$^{-2}$ &&  & EW & $\sim$ & 240~eV\\\hline

\end{tabular}
\end{center}
\label{mcg}
\end{table*}

\subsection{Broad Fe K line}

It is well known that the shape and strength of the broad component of 
the Fe K line, so
important as a diagnostic of the innermost regions close to the black
hole (\eg Fabian \et 2000), depends critically on fitting the underlying 
continuum correctly.
In the examples shown in Figure 1 it seems possible to 
identify the
power law component in all cases, except perhaps MCG -6-30-15.
In Mkn 766, where a broad Fe K line is widely
reported (Nandra \et 1997, Leighly 1999, Mason \et 2002), visual examination 
of the spectral ratio does indicate such a
feature. One cautionary remark, however, is to point out the 
importance of correctly fitting the data above $\sim7$~keV. This is not well 
constrained by \xmm data (and was even less so in \asca), but there is some 
visual indication of
absorption above the Fe K edge (Figure 4(a)). Such absorption could be a 
consequence of
`reflection', or may arise in \los matter, neither of which are as yet 
very well constrained. It is interesting to note that the inclusion
of an absorption edge in the fit to the EPIC data of Mkn 766 (Figure 4(b)) 
has a 
significant effect on the parameters of the broad Fe K line, as it
lowers 
the underlying power law index by $\sim0.05$.

Ignoring an absorption edge, the underlying power law fitted to the data at 
2--5 keV and above 7 keV (the typical \asca recipe) is $\Gamma$~$\sim2.01$. 
A broad line is
required to fit the residuals and a good fit ($\chi^{2}$/dof of
733/794) is obtained with a Kerr
line at 6.5 keV with EW of $\sim450$~eV. Inclusion of an edge at
$\sim7$~keV and fitting the power law between 2--6 keV and 6.8--12 keV 
reduces the power law slope to $\Gamma$~$\sim1.95$. No broad line is
then required, a good fit over the 2--12 keV band being obtained
($\chi^{2}$/dof of 723/795) with just the
absorption edge ($\sim7.1$~keV, $\tau \sim0.17$) and a relatively narrow Fe K 
line (E~$\sim6.55$~keV, $\sigma\sim0.18$~keV, EW $\sim122$~eV).
We note that from the \xmm data $\it alone$ (and implicitly also for
\asca data) the absorption edge provides the main determinant of
reflection, which in this case is consistent with an origin in the
`cold, distant matter' responsible for the narrow Fe K line.  

\begin{figure}
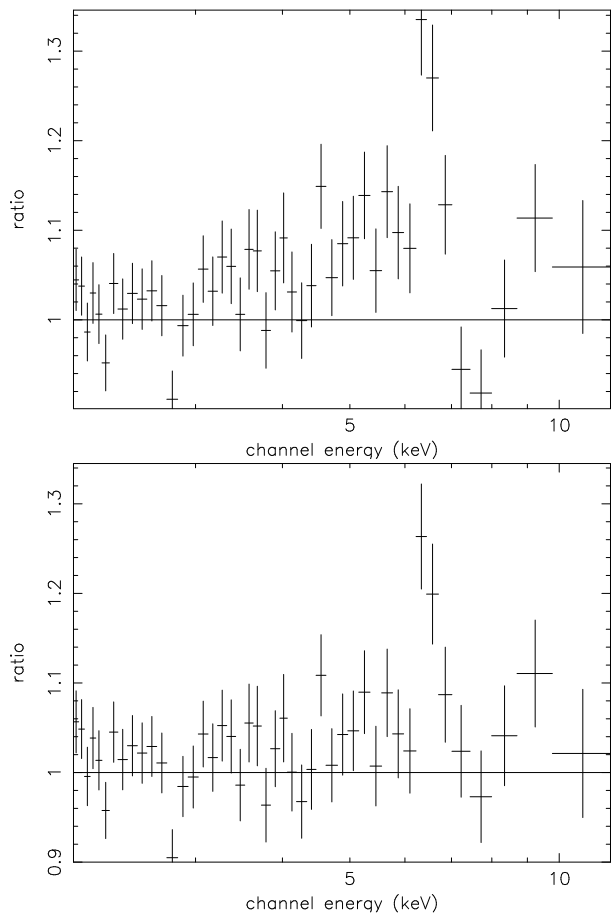

\centering
\begin{minipage}{85 mm} 
\centering
\vbox{
\includegraphics[width=6 cm, angle=270]{kpounds-C2_fig4a.ps}
\includegraphics[width=6 cm, angle=270]{kpounds-C2_fig4b.ps}
}
\caption{Data/model ratio plots for (a) a power law and (b) a power
law plus absorption edge fit to the Mkn 766 data}
\end{minipage}
\end{figure}

\begin{figure}
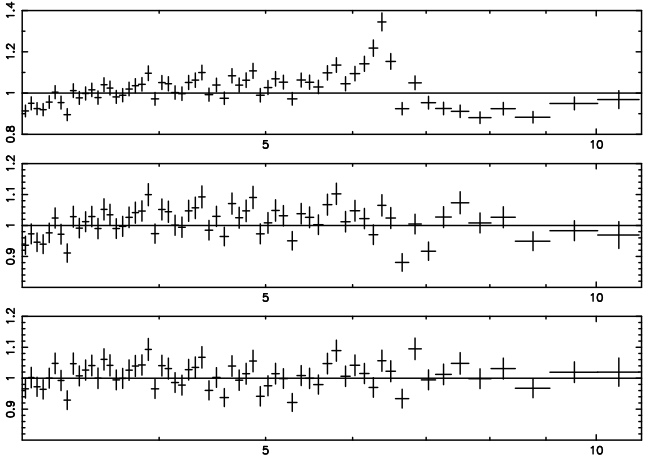

\centering
\begin{minipage}{85 mm} 
\centering
\vbox{
\includegraphics[width=2 cm, angle=270]{kpounds-C2_fig5a.ps}
\includegraphics[width=2 cm, angle=270]{kpounds-C2_fig5b.ps}
\includegraphics[width=2 cm, angle=270]{kpounds-C2_fig5c.ps}
}
\caption{Data/model ratio plots for fits to (a) a power law, (b) a power plus 
narrow Fe K line and 
absorption edge, (c) a power law, absorption edge plus narrow and broad
Fe K lines fit to the MCG-6-30-15 EPIC data}
\end{minipage}
\end{figure}

It is interesting, finally, to test the `prototype' broad Fe K line
Seyfert, MCG-6-30-15 (Tanaka \et 1995, Iwasawa \et 1996, Wilms \et 2001) 
with a model that allows for the Fe K absorption
edge indicated in the \xmm data. Figure 5(a) shows the deviations from
a simple power law fit over the 3--12 keV band. 
As with Mkn 766, an acceptable spectral fit is obtained over this band 
($\chi^{2}$/dof of 557/494)
with a power law, narrow Fe K line and absorption edge. The best-fit
parameters are:
a power law of $\Gamma$~$\sim1.74$; a narrow line at $\sim6.37$~keV,
with width $\sigma\sim0.09$~keV, and EW $\sim74$~eV; 
and an edge of depth $\tau\sim0.24$ at 
$\sim7.25$~keV (Figure 5(b)).
However, a still better fit is obtained ($\chi^{2}$/dof of 511/488) with the
addition of an extreme broad
component to the Fe K line, at $\sim6.33$~keV, with disc emissivity 
$\beta$ of $\sim4.7$ and
EW of $\sim240$~eV, together with a narrow line
($\sim6.39$~keV, sigma $\sim0.11$~keV, EW $\sim95$~eV) and 
absorption edge ($\sim7.32$~keV, $\tau \sim0.11$). The power law slope 
is essentially unchanged at $\Gamma$~$\sim1.75$. Examination of the residuals 
of the previous fit (Figure 5 (b))
suggests the broad line is modelling quite subtle curvature in the EPIC data
between 3--6 keV. 

Extension of 
the latter model over the full EPIC spectrum (0.5--10 keV) also
gives a remarkably good fit, where the spectral curvature below
$\sim3$~keV is modelled by the addition of line-of-sight ionised gas 
with a column 
density $\sim10^{22}$ cm$^{-2}$ and ionisation parameter $\xi\sim5.9$ 
(Figure 6).
Clearly this fit (summarised in Table 2) is only a crude approximation
to the complex spectrum
seen in the RGS and \chandra data; however, it is reassuring that an
acceptable broad-band fit can be made to this most complex spectrum (see
Figure 1), offering the promise of developing self-consistent models of
Seyfert X-ray spectra in which the primary X-ray continuum can be
resolved from the superimposed absorption and emission features.

\begin{figure*}
\begin{minipage}{170 mm} 
\includegraphics[width=12 cm, angle=270]{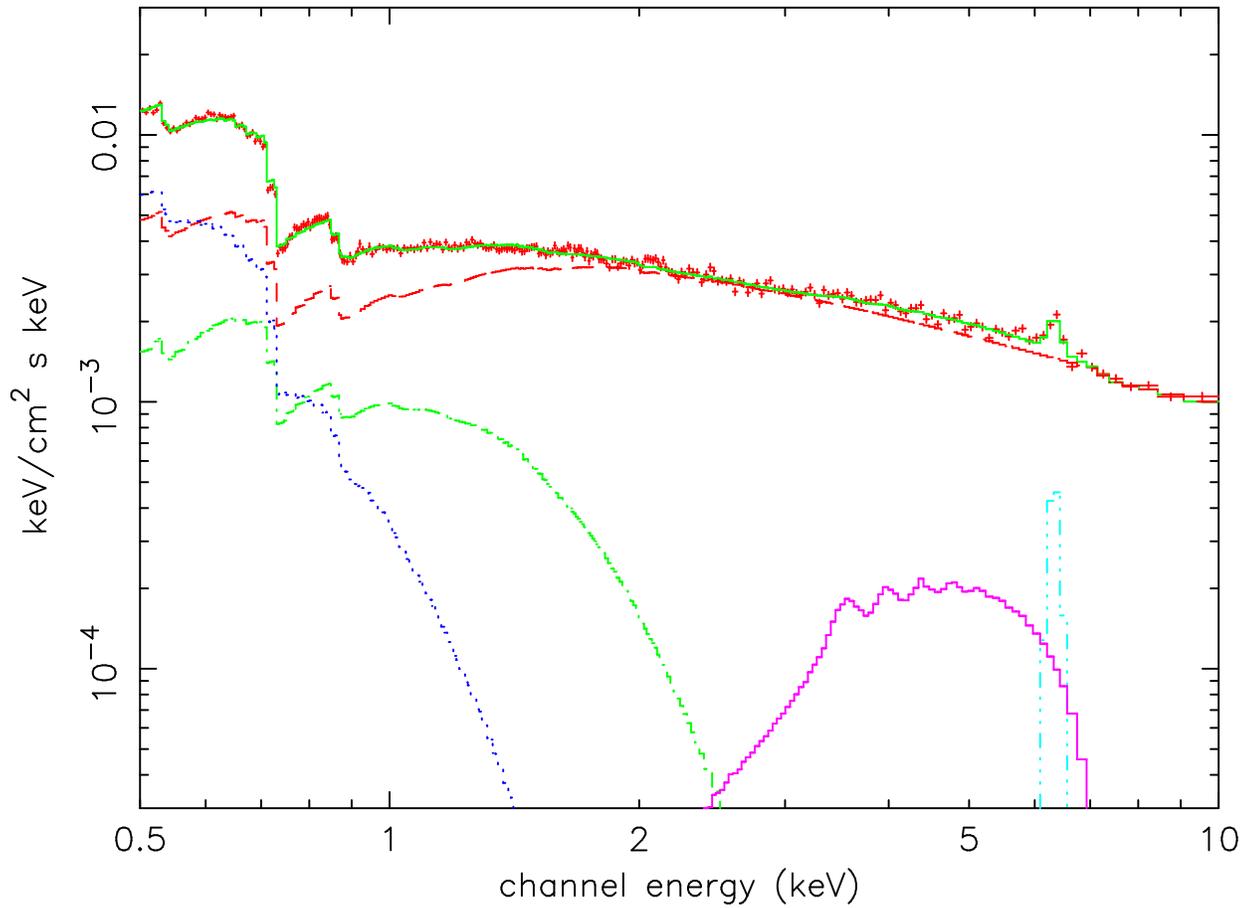}
\caption{An unfolded spectral plot of the EPIC data on MCG-6-30-15 for
a model consisting of: a power
law plus broad (Kerr) and narrow Fe K lines; an Fe K absorption
edge; soft excess (blackbody) emission; 
plus low energy absorption by ionised gas.}
\label{MCG-6}
\end{minipage}
\end{figure*}

\section{Summary}

The most striking aspect of the range of EPIC spectra
displayed in Figure 1
is the presence in every case of a strong
soft emission component which causes the X-ray continuum to turn up sharply
below $\sim$~0.7 keV in the lower luminosity sources (SSX), and is evident 
as a gradual up-turn (GSX), below $\sim$~3 keV for the high luminosity
sources.
The form and location of the SSX is strongly suggestive of a link with 
K-shell oxygen, though - unsurprisingly - the lower resolution EPIC 
data can not distinguish between the alternative descriptions
(relativistically broadened line emission, Sako \et 2001; complex
absorption, Lee \et 2001) currently
being advocated for MCG-6-30-15 and Mkn 766.
RGS spectra do confirm
the absence of strong absorption in the higher luminosity sources in
our sample, Mkn 509 and PKS 0558-504. This leads us to speculate that
a broad soft emission component extending up to $\sim$~3~keV (arising by
Comptonisation of cool disc photons) may be
common to our sample, but is obscured by increasingly strong absorption 
from ionised 
gas for the lower luminosity Seyferts 
NGC 5548, Mkn 766 and MCG-6-30-15. We emphasise the complementary strengths of
broad-band data from EPIC and higher resolution spectra from the RGS
in providing the input for improved modelling of the primary X-ray
continuum and - consequently - of absorption and secondary emission
features that carry unique information on the X-ray emission mechanism
and nuclear environment in Seyfert galaxies.

One such secondary emission feature, the narrow Fe K emission line at
6.4 keV, is found to be
a common property across our sample (apart from the highest luminosity 
object PKS 0558-504). The EW of $\sim$~50--100 eV is
consistent with reflection from cold, distant matter subtending 
a solid angle of 1--2 $\pi$ steradians at the hard X-ray source. Line
profile and variability studies offer the promise of using this
emission line to probe dense cold matter from the outer disc to the 
molecular torus.

We suggest the broad Fe K line may not be as common, or as strong, 
as it appeared to 
be from earlier 
\asca
observations. A key factor is the need to correctly model the spectrum
above $\sim$~7 keV, where \xmm has improved - though still limited - 
sensitivity. Also, as spectral fits to the MCG-6-30-15 data show,
precise calibration of X-ray optics and detectors is critical to
validating measurements of the important broad Fe K line.

\begin{acknowledgements}

We thank Rick Edelson
for permission to work on the EPIC data from his GO observation of NGC
5548 and to
Jorn Wilms for access to the GT observation of MCG-6-30-15.  
Grateful thanks are also due to the dedicated efforts of the 
many colleagues who delivered and
continue to sustain the powerful \xmm Observatory. \xmm is funded
by the European Space Agency with support from NASA. The EPIC 
instrument was funded by the UK, Germany, France and Italy and developed 
by a team
led by Martin Turner.

\end{acknowledgements}

\end{document}